# Electromagnetic acceleration of electrically charged bodies

S. N. Dolya

*Joint Institute for Nuclear Research, Joliot Curie str. 6, Dubna, Russia, 141980*

**Abstract**

Acceleration of electrically charged bodies is carried out by the electric field running via the spiral structure of the electric pulse. The accelerated particles have a cylindrical shape with a diameter of cylinder $d_{sh} = 2$ mm, a length of the conical part $l_{cone} = 13$ mm and the total length $l_{sh} = 300$ mm. Pre-acceleration of the cylinder up to speed $V_{in} = 1$ km / s is performed by gas-dynamic. The pulse with the voltage amplitude $U_{acc} = 2$ MV and the power P = 300 MW goes into the spiral waveguide synchronously with the rod injected onto it. The rod is accelerated by the traveling pulse in the longitudinal direction up to the finite velocity $V_{fin} = 6$ km / s for length $L_{acc} = 300$ m.

**Introduction**

There is a known [1] method of magnetic dipole acceleration by sequential current turns, which are switched on one by one creating the current pulse moving in the space. Magnetic dipoles in this process are accelerated by the magnetic field gradient in the space. In principle, using a large number of turns allows one to achieve a high finite speed of the magnetic dipoles.

Such a multiple-section accelerator consisting of a sequence of coils and capacitors can be considered as a line with the lumped parameters. If you move from the line with lumped parameters to the line with distributed parameters, you will obtain a casual coaxial cable, which has the central wire rolled into a spiral or on a spiral waveguide.

In such a cable in a wide range of frequencies there is no dispersion, i.e. there is no dependence of the phase velocity on the wave frequency. The phase velocity in this cable coincides with the group velocity of the wave. The wave propagation velocity (pulse), V, in such a cable is defined by the tightness of winding of the central conductor in a spiral as well as by the dielectric properties of the medium which fills the cable. This ratio is called the dispersion equation and looks as follows:

$$\beta = \mathrm{tg}\Psi/\varepsilon^{1/2}, \qquad (1)$$

$\beta = V / c$, V - velocity of the pulse propagation via the cable,

$c = 3 *10^{10}$ cm / s - the speed of propagation of electromagnetic waves in the vacuum, tg $\Psi = h/2\pi r_0$, h -the winding step of the spiral, $r_0$ - the radius of the spiral winding, ε - relative dielectric constant of the medium filling the cable. The wave as if runs over the circle spiral around $2\pi r_0$ while moving at a small distance h along the axis of the spiral. Further the wave additionally slows down due to the dielectric properties of the medium defined by the value of ε.

To speed up the rod by the pulse running in the cable, the area inside the spiral must be released from the dielectric. Thus, the speed of the pulse in this cable will be slightly increased, [2]:

$$\beta = \sqrt{2} * \text{tg } \Psi/\varepsilon^{1/2}. \qquad (2)$$

Running on the line with lumped or distributed parameters the pulse contains not only the gradient of the magnetic field which accelerates the magnetic dipoles, but also the electric field $E_{zw}$, which can accelerate charged bodies.

**1. The parameters of the body being accelerated**

We will consider the acceleration of the rod with a conical head which is electrically charged.

Acceleration of macro particles in a spiral waveguide is well-known, [2]. For this acceleration it is required that the initial velocity of the rod and the phase velocity are approximately the same. When we accelerate the rod, the phase velocity in the spiral waveguide should be increased for the rod to be all the time in one the same phase of the wave, which is called a synchronous phase. To increase the phase velocity of the wave in the waveguide is possible by increasing the winding step of the spiral or decreasing its radius, or doing the both simultaneously, [2].

Let the diameter of the rod be equal to: $d_{sh} = 2$ mm, length $l_{sh} = 300$ mm. Then the cross-section area of the rod is: $S_{tr} = \pi d_{sh}^2 / 4 = 3.14 * 10^{-2}$ cm$^2$, the volume of the rod: $V_{sh} \approx 1$ cm$^3$. The weight of the rod in the case of the average density of the rod $\rho_{aver} = 5$ g/cm$^3$, is equal to $m_{sh} = 5$ g.

*1.1. The ratio Z / A*

We assume the average atomic mass of the rod to be equal to $A_{sh} = 30$. We find the number of nucleons in the rod from the following proportion:

$$6 * 10^{23} \text{ -------- } 30 \text{ g}$$
$$x \text{ ---------- } 5 \text{ g,}$$

where $x = 10^{23}$ atoms or $A_{sh} = 3 * 10^{24}$ nucleons.

We take the surface tension of the electric field on the rod to be: $E_{surf} = 3 * 10^7$ V / cm. The formula for the surface tension of the electric field for the cylinder is the following:

$$E_{surf} = 2\kappa / r, \qquad (3)$$

we find the charge density per the length unit of the rod:

$$\kappa = E_{surf}\, r/2e = (3 * 10^7 * 0.1) / (5 * 10^{-10} * 300 * 2) = 10^{13}, \qquad (4)$$

from where you can find

$$N_e = (\kappa / e) * l_{sh} = 3 * 10^{14}. \qquad (5)$$

Thus, if the "put" $N_e = 3 * 10^{14}$ electrons on the rod, the surface tension of the field will turn out to be: $E_{surf} = 3 * 10^7$ V / cm.

Now when we know the total number of the electrons "placed" on the rod: $N_e = 3 * 10^{14}$, and the number of nucleons on it: $A_{sh} = 3 * 10^{24}$, it is possible to find the ratio of the charge to the mass for the rod: $Z / A = N_e / A =$
$= 3 * 10^{14} / 3 * 10^{24} = 10^{-10}$.

*1. 2. Irradiation of the rod with the electron beam*

To accelerate a cylindrical rod with a cone head in a spiral waveguide, it is necessary to charge it electrically. To give the electric charge to the rod is possible by irradiating it with the electron beam; so that the electrons irradiating the rod would obligatory remain on it. Then the electric charge of the rod will increase proportionally to the current of the electron beam and the duration of exposure of the rod. Suppose that the current of the electron beam irradiating the rod is equal to $I_{beam} = 5$ A, and the current pulse duration is $\tau_{beam} = 10$ μs. Then the total number of electrons in the current pulse is exactly equal to $N_e = I_{beam} * \tau_{beam} / e = 3 * 10^{14}$ electrons.

### 1.3. Irradiation of the rod with the electron beam. The electron energy

Let the cylindrical rod accelerated by the gas-dynamic method up to speed $V_{in} = 1$ km / s be irradiated with the electron beam obtained from an external source. We assume the surface tension of the electric field to be equal to $E_{surf} = 30$ MV / cm. Then, for the diameter of the cylinder $d_{sh} = 2$ mm, we find that the minimum energy of the electrons which can overcome the Coulomb repulsion of the electrons previously placed on the rod should be:
$W_e > eE_{surf} * d_{sh} / 2 = 3$ MeV.

### 1. 4. Irradiation of the rod with the electron beam.
### The path length of the electrons in the rod

Electrons with the energy of 3 MeV have the range path in aluminum approximately equal to 1 g/cm$^2$, [3], page 957. Assuming the density of aluminum to be equal to: $\rho_{Al} = 2.7$ g/cm$^3$, we find that the extrapolated path of electrons in aluminum is: $l_{Al} \approx 4$ mm. Since the average density of the material chosen for the cylinder $\rho_{aver} = 5$ g/cm$^3$, that is by about twice more than the density of aluminum, the path length of electrons of the energy of 3 MeV in the rod will be approximately equal to 2 mm.

Evidently, it is necessary to gradually increase the energy of the electrons in the process of irradiation. It is needed that while "setting" the electrons on the rod, the electrons emitted later, on the one hand, would have a sufficiently high energy to overcome the Coulomb repulsion of the electrons being already on the rod, and, on the other hand, the electron energy must not be too high because it is necessary to have the length of the electron path in the material of the rod to be much less than its diameter.

In this energy range the length of the electron passing in the rod material linearly increases with the energy, for example, of the electrons with energies $W_e = 300$ keV, having the length of electron passing equal to 0.2 mm. They cannot cross the rod diameter of 2 mm. They will lose their energy for ionization of the matter, and will be placed on the rod.

## 1. 5. Irradiation of the rod with the electron beam. Field electron emission

To "plant" several charges on the rod is not a problem, but when there are many electrons on the rod, they will start to leak out from it due to the field emission. Let the field strength for the field emission be $E_{surf} = 3 * 10^7$ V / cm. When there are enough "planted" electrons on the rod, to plant the next portion, it is necessary to overcome the repulsion of those electrons which are already sitting there. This means that the energy of the electrons, which we would like to put on the rod, should be large enough so that they can overcome this Coulomb barrier, reach the rod and stay on it.

"Planting" a large electric charge would be interfered by the field emission. A part of the charge due to the tunnel effect will continuously leak out from the rod.

## 1.6. Surface covering with platinum and oxygen passivation of the cylinder

To create a surface barrier for the electrons "having placed" on the rod, it is needed to increase the energy yield of the electrons from the rod. The largest energy yield belongs to platinum passivated by oxygen, $e\varphi = 6.56$ eV, [3], page 445. Planted on the rod the charge will leak out from it by the field emission according to the formula [3], page 444:

$$j = e^2E^2/(8\pi h\varphi)*\exp\{ [-(8\pi/3)(2m)^{1/2}/h]*[(e\varphi)^{3/2}/(eE)*\theta(y)]\}, \quad (6)$$

where $\theta(y)$ is the Nordheim function. The argument of this function is a relative reduction of the energy yield by the external electric field according to Schottky's law.

## 1.7. Leakage of electrons

Let us find the number of electrons which will leave the rod during acceleration. For the field tension $E = 30$ MV / cm and the energy yield $e\varphi = 6.5$ eV from the graph, [3], page 461, we find that the leakage current density is: $j = 10^{-9}$ A/cm$^2$.

Leakage of charge $\Delta Q$ will be:

$$\Delta Q = j * S_{surf} * t_{acc}, \quad (7)$$

where j = $10^{-9}$ A/cm$^2$ - leakage density current, $S_{surf} \approx 20$ cm$^2$ – the total surface of the rod.

The acceleration can be determined from the following formula:

$$t_{acc} = L_{acc} / V_{aver}, \qquad (8)$$

where $L_{acc}$ = 300 m - length of acceleration, $V_{aver}$ = 3 km / s - the average speed over the length of the acceleration. Calculating the time of the acceleration from formula (8) we find it to be equal to: $t_{acc}$ = 0.1 s.

Substituting numbers into the formula (7) we find that $\Delta N_e = 10^{10}$ electrons and it is 3 * $10^{-5}$ - the number of electrons, which were planted on the rod.

**2. The acceleration length**

The acceleration rate of the charge in the electric field can be written as follows:

$$\Delta W = (Z / A) \, eE_{zw}, \qquad (9)$$

and, for the tension of the wave $E_{zw}$ = 70 kV / cm, the rate of the energy gain will be: $\Delta W = 7 * 10^{-4}$ eV / (m * nucleon), so that the required increase of energy $W_{fin}$ = 0.2 eV / nucleon will be reached on the length:

$$L_{acc} = W_{fin} / \Delta W = 300 \text{ m}. \qquad (10)$$

**3. Selection of parameters of the spiral waveguide**

The initial velocity of the rod in a spiral $\beta_{sh\ in}$ expressed in terms of the velocity of light $\beta_{sh\ in} = V_{sh\ in} / c$, where c = 3 * $10^{10}$ cm / s, the velocity of light in vacuum is equal to $\beta_{sh\ in}$ = 3.3 * $10^{-6}$, finite $\beta_{sh\ fin}$ = 2 * $10^{-5}$. The spiral is assumed to consist of several sections, so that within each section to select the optimal acceleration rate. The wavelength of the acceleration can be determined from the condition: x = $2\pi r_0 / (\beta_{ph} * \lambda_0) = 1$, where x - a dimensionless parameter which is the argument of the modified Bessel functions, $r_0$ - the radius of the spiral, $\beta_{ph}$ - phase velocity, $\lambda_0$ - wavelength acceleration in the vacuum, $\lambda_0 = c/f_0$, $f_0$ – acceleration frequency .

Choosing the initial radius r0 in the spiral equal to $r_{0\,in} = 20$ cm, $\varepsilon = 1280$ - the dielectric constant of the medium located in the area between the coil and the screen, we find: $\lambda_0 = 3.8 * 10^7$ cm, $f_0 = 790$ Hz. Thus, the slowdown wavelength for the start of acceleration is equal to: $\lambda_{slow} = \beta\lambda_0 = 1.25$ m.

*3.1. Parameters of the spiral*

In order to obtain the required field intensity $E_0$ in a spiral waveguide, it is required to introduce the power, defined by the formula, [2]:

$$P = (c / 8) * E_0^2 * r_0^2 * \beta_{ph} * \{\}, \qquad (11)$$

where P – the high frequency power introduced into the spiral waveguide, $r_0$ - the radius of the spiral, $\beta_{ph}$ - phase velocity of the wave, which is determined from the dispersion equation. The curly bracket in (11) is equal to:

$$\{\} = \{(1+I_0K_1/I_1K_0)(I_1^2-I_0I_2) + \varepsilon\,(I_0/K_0)^2(1+I_1K_0/I_0K_1)(K_0K_2-K_1^2)\}, \quad (12)$$

where $I_0$, $I_1$, $I_2$ are the modified Bessel functions of the first kind, $K_0$, $K_1$, $K_2$ - the modified Bessel functions of the second kind. The first term in the curly bracket corresponds to the flux propagating inside the spiral; the second term corresponds to the flux traveling outside the spiral. Since the space between the spiral and the screen is filled with a dielectric, before it there is the second term factor $\varepsilon$, [2].

In this case, to slow down the electromagnetic wave till the velocity of sound, it is required to use geometrical properties of the structure (spiral small step) as well as the properties of the medium. That is why we have chosen the relative permittivity $\varepsilon = 1280$.

Thus, the flow of high frequency power propagating outside the spiral is more than $10^3$ times greater than the power propagating inside the spiral. Therefore, the first term inside the curly bracket can be neglected compared to the second one. The very meaning of the bracket for the argument $x = 1$ is approximately equal to: $\{\} \approx 4\varepsilon$.

In accelerators the synchronous phase is selected on the front slope of the pulse, so that the electric field accelerating the particle is always less than the amplitude value. Let us choose a synchronous phase to be equal to: $\varphi_s = 45^0$, $\sin\varphi_s = 0.7$, $E_{zw} = E_0\sin\varphi_s$. Thus, the amplitude of the wave which accelerates the cylindrical rod should be equal to the following:

$$E_0 = E_{zw} / \sin\varphi_s = 100 \text{ kV} / \text{cm}. \qquad (13)$$

Then, the wave power, expressed by the formula (11) in Watts, is equal to:

$$P \text{ (W)} = 3*10^{10}*10^{10}*4*10^2*3.3*10^{-6}*1.28*10^3*4/(8*9*10^4*10^7) =$$
$$= 300 \text{ MW}. \qquad (14)$$

*3.2. The transition from a sine wave to a single pulse*

This power can be achieved by using the pulse technique. We expand the sinusoidal pulse, [2], the corresponding half-wave $E_{pulse} = E_{0pulse}\sin(2\pi/T_0) t$, $2\pi/T_0 = \omega_0$, $\omega_0 = 2\pi f_0$ in a Fourier series:

$$f_1(\omega) = (2/\pi)^{1/2} \int_0^{T_0/2} \sin\omega_0 t * \sin\omega t dt. \qquad (15)$$

The pulse spectrum is narrow and covers the frequency range from 0 to $2\omega_0$. Since the spiral waveguide dispersion (dependence of the phase velocity on the frequency) is weak, it can be expected that the full range of frequencies from 0 to $2\omega_0$ will propagate approximately with the same phase velocity. As a result, the half-wave sinusoidal pulse will spread out in the space and becomes wider by several times only due to the increase of the phase velocity of the wave. In this case the spiral waveguide is necessary to match with a supply feeder in the following frequency range: $\Delta f \approx \omega_0/2\pi$.

We introduce the concept of pulse amplitude $\tilde{U}$, associated with the field tension at the axis of the spiral $E_0$ by the following ratio, [2]:

$$\tilde{U}_{pulse} = E_{0pulse}\lambda_{slow}/2\pi, \ \lambda_{slow} = \beta\lambda_0, \ \lambda_0 = c/f_0. \qquad (16)$$

Selection of wavelength $\lambda_0 = 3.8 * 10^7$ cm means that we have chosen the duration of the acceleration of the rod equal to ($f_0 = c/\lambda_0 = 790$ Hz), $\tau_{pulse} = 1/(2f_0) = 630$ µs. The amplitude of the voltage pulse will be equal to: $\tilde{U} = E_0\lambda_{slow}/2\pi = 2$ MV. Table 1 summarizes the main parameters of the accelerator.

Table1. The parameters of the accelerator

| | |
|---|---|
| $Z/A = 10^{-10}$, insulator outside the spiral, wave power, P | P = 300 MW $\mu = 1$, $\varepsilon = 1280$ |
| Speed, the initial - finite, $\beta_{ph}$ | $\beta_{ph} = 3.3*10^{-6} - 2*10^{-5}$ |
| Initial radius of the spiral, $r_0$ | $r_0 = 20$ cm |
| Frequency of the wave, $f_0$, | $f_0 = 790$ Hz |
| Tension of the electric field $E_0$ | $E_0 = 100$ kV/cm |
| Length of the accelerator, $L_{acc}$ | $L_{acc} = 300$ m |
| Pulse duration, $\tau$ | $\tau = 630$ μs |
| Amplitude of voltage $\tilde{U}_a$ | $\tilde{U}_a = 2$ MV |

*3. 3. The capture of particles in the acceleration mode. Admission*

We calculate the required accuracy to match the initial phase of the accelerating wave (pulse) with a synchronous phase. The theory of particle capture in a traveling wave gives $\Delta\varphi = 3\varphi_s$, $(+ \varphi_s - 2\varphi_s)$, [4]. In our case it means the following: $T_0 / 4$ correspond to the duration of 316 μs or $90^0$ degrees, and a one-degree phase corresponds to the time interval of approximately 3 μs. In linear accelerators the buncher gives the phase width of the bunch $\pm 15^0$. To avoid large phase oscillations, it is required that the timing accuracy of synchronization of the rod with the accelerating pulse would be equal to: $\Delta\tau = \pm 15 * 3$ μs $= \pm 45$ μs. This timing precision seems to be quite achievable for the gunpowder start which is a preliminary gas-dynamic acceleration of the rod.

Now let us calculate the required accuracy of the coincidence of the wave phase velocity with the initial rate of the rod. We introduce value $g = (p-p_s) / p_s$ - the relative difference between the pulses, [4]. In the non-relativistic case – it is just the relative velocity dispersion of $g = (V-V_s) / V_s$. The vertical scale of the separatrix is calculated by the following formula, [4]:

$$g_{max} = \pm 2 [(W_\lambda ctg\varphi_s/2\pi\beta_s) * (1 - \varphi_s / ctg\varphi_s)]^{1/2}, \qquad (17)$$

wherein: $\varphi_s = 45^0 = \pi / 4$, $ctg\varphi_s = 1$, $[1 - \varphi_s / ctg\varphi_s]^{1/2} = 0.46$, $2 * 0.46 = 0.9$
$W_\lambda = (Z / A) eE_0\lambda_0 \sin\varphi_s/Mc^2$.

Let us determine the value of $W_\lambda = (Z / A) * eE_0\lambda_0 \sin\varphi_s/Mc^2$, which is the relative set rod energy at wavelength $\lambda_0$ in vacuum. In our case

$\lambda_0 = c/f_0 = 3.8 * 10^7$ cm, $\sin\varphi_s = 0.7$, $Mc^2 = 1$ GeV, $W_\lambda = 2.66 * 10^{-6}$.
Substituting numerical values, we get $g = (V_{in}-V_s) / V_s = \Delta V / V_s$, and, finally,

$$\Delta V / V_s = \pm [2.66 * 10^{-6} / (6.28 * 3.3 * 10^{-6})]^{1/2} * 0.9 = \pm 0.11.$$

Thus, the accessible mismatch of the rod initial velocity with the pulse velocity is of the order of $\Delta V / V_s = \pm 11\%$. For the initial rate of the rod $V_{in} = 1$ km / s, the mismatch accuracy of the velocity deviation is $\Delta V < 100$ m / s.

**4. Radial movement**

It is well known, [4], that in the azimuthal - symmetric wave the phase stability region corresponds to the radial defocusing. In this case, you cannot simultaneously obtain both the radial and phase stabilities. Under the conditions of phase stability for radial focusing it is required to use the external field. In this phase region the radial component of the electric field of the wave is directed to the increasing radius, i.e. it radially accelerates the rod.

In this velocity region of the rods – the "hypersonic" velocity region where they are by hundreds of thousand times smaller than the speed of light, the focusing by magnetic quadrupole lenses is not efficient. In this case the most suitable focusing is by using the electrostatic quadrupole lenses. These lenses focus the particles in one plane and defocus them in the other one. Collected into a doublet, two lenses of this sort give the focusing effect. The accelerator is divided into separate sections and the focusing doublets are placed between the accelerator sections.

## 5. Operation of the device

Fig. 1 shows a diagram of the apparatus.

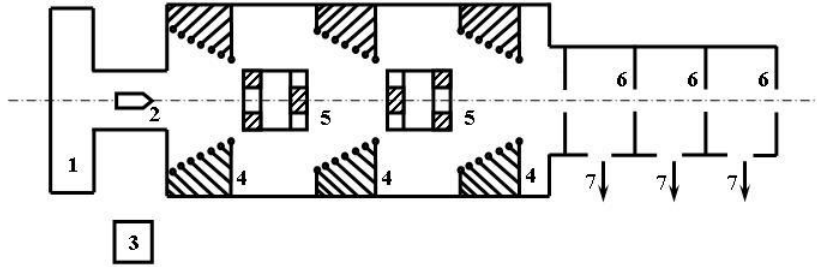

Fig. 1

The apparatus operates as follows. Inside the barrel there is the gun – (1), (2) - a cylindrical rod with a sharp conic head, which is accelerated till the speed corresponding to the speed of injection in a spiral waveguide till $V_{in} = 1$ km / s. From the linear accelerator (3) to the rod which is irradiated by the beam of electrons with energy $eE = 3$ MeV, the total number of electrons is planted on the rod $N_e = 3 * 10^{14}$. Therefore, the electric field tension obtained at the surface of the cylinder is equal to $E_{surf} = 3 * 10^7$ V / cm, the potential of the cylinder - $e\Phi = 3$ MeV, the ratio of the planted charge to the mass is $Z / A = 10^{-10}$. The electric field potential of the high current pulse with voltage $\tilde{U}_a = 2$ MV, is propagating via sections (4) of the spiral waveguide of the total length of $L_{acc} = 300$ meters. The rod is accelerated to the finite speed $V_{fin} = 6$ km / s. The electrostatic quadrupole lens doublets (5) are placed between the sections and they focus the rods in the transverse direction. The rods are released into the atmosphere through a series of buffer volumes (6). Each buffer volume has individual pumping (7).

**Conclusion**

The flight parameters of the rod are represented in the Table 2, where they are given as a function of time in the first column. The second column shows the vertical velocity of the rod, the third one - the horizontal velocity of the rod, the fourth column shows the achieved altitude of the rod, the fifth one represents the density of the atmosphere at this altitude.

Table 2. The flight parameters depending on the time, for the case of $C_x, C_y = 2.5*10^{-2}$.

| t, s | $V_x$, km/s | $V_y$, km/s | Y, km | $\rho_{air}$, g/cm$^3$ |
|---|---|---|---|---|
| 0 | 6 | 0 | 0 | $1.3*10^{-3}$ |
| 10 | 3.72 | 3.67 | 18 | $6*10^{-5}$ |

The time of flight up to the maximum altitude is equal to
$\tau_{max} = V_y / g = 367$ s, where $g = 10^{-2}$ km/s$^2$ (gravity acceleration range). The distance of the flight is $S = V_x * 2\tau_{max} = 2700$ km, the maximum altitude is $Y = V^2_y/2g = 670$ km.